\documentclass[reprint,showpacs,aps,pra,amsmath,amssymb,floatfix]{revtex4-1}
\usepackage{graphicx,epstopdf}
\usepackage{bm}
\usepackage{color}
\usepackage[bookmarks,colorlinks,linkcolor=blue,anchorcolor=black,urlcolor=blue,citecolor=blue]{hyperref}
\usepackage{natbib}
\usepackage{CJK}

\allowdisplaybreaks[3]

\begin{document}
\begin{CJK*}{UTF8}{gbsn}

\title{Hidden PT Symmetry and quantization of coupled-oscillators model of QASER}

\author{Lida Zhang$^{1}$~(\CJKfamily{gbsn}张理达)}

\author{G. S. Agarwal$^{1}$}

\author{W. P. Schleich$^{1,2}$}

\author{M. O. Scully$^{1}$}

\affiliation{$^{1}$ Texas A \& M University, College Station, Texas 77843, $^{2}$ Universit\"{a}t Ulm, D-89081 Ulm, Germany}

\date{\today}

\begin{abstract}
Using Maxwell-Bloch equations it has been shown how the superradiance can lead to amplification and gain at a frequency much larger than the pumping frequency. This remarkable effect has been examined in terms of a simpler model involving two coupled oscillators with one of them parametrically driven. We show that this coupled oscillator model has a hidden parity-time (PT) symmetry for QASER, we thus bring PT symmetry to the realm of parametrically coupled resonators. Moreover, we find that the QASER gain arises from the broken PT symmetry phase. We then quantize the simplified version of the QASER using quantum Langevin equations. The quantum description enables us to understand how the system starts from quantum fluctuations.

\end{abstract}







\maketitle

\end{CJK*}

\section{Introduction}

Since the advent of lasers at the last 60s, it has not only pushed science forward in almost every aspect but also altered humans' life to an unprecedented level. As laser and laser-based technologies have become one of the indispensable building blocks of our modern society, it is natural to think of higher-frequency lasers in X-ray regime which can provide untrahigh intensity and precision (e.g., nonoscale) compared to current optical technology, 
and have great potential in coherent imaging of macromolecule structures and tissues which are crucial in medicine, chemistry and biology~\cite{Schroer2008PRL,Robinson2009NMat,Chapman2010NPhoton}. Unfortunately, owing to the very short wavelength of X-ray which can penetrate normal mirrors, the X-ray laser can not be simply generated using active cavities as optical lasers. 
It is therefore vital to find new physical mechanisms to generate X-ray laser in a compact and efficient way. 

An outstanding and promising example is Quantum Amplification by Superradiant Emission of Radiation (QASER)~\cite{Svidzinsky2013PRX,Scully2014LP}, which utilizes an intense low-frequency laser to drive the atomic ensemble around its lower collective frequency (infrared) which matches the frequency difference between the two normal modes with much higher frequencies (X-ray) in the light-atom interacting system, resulting in amplification at the higher-frequency. The essence of QASER is coupling atomic superradiance and a combination parametric resonance to get exponential generation of X-rays. Surprisingly the QASER shares many aspects of two-coupled oscillators asymmetrically and parametrically driven by a low-frequency field~\cite{Svidzinsky2013PRX,Chen2016PS}. The asymmetry in the parametric coupling is especially interesting as in physics typically we deal with Hermitian couplings. We show in this paper that this very asymmetric coupling is associated with the hidden parity-time (PT) symmetry in the model.


We note that the study of the PT-symmetric Hamiltonians~\cite{Bender1998PRL,Bender2007RPP} has attracting considerable attentions especially after experimental realization of the PT-symmetry in optical systems~\cite{Guo2009PRL,Ruter2010NPhys,Regensburger2012Nature,Peng2014NPhys,Chang2014Nphoton,Liu2016PRL,He2016PRA,Xu2015PRA}. 
The most prominent feature of PT symmetry is that PT-symmetric Hamiltonians can give either real or complex eigenvalues depending on the ratio between its real and imaginary parts.
If all the eigenvalues of the PT-symmetric Hamiltonian are real, the system is said to be in an unbroken PT-symmetry phase; Otherwise, it refers to a broken PT-symmetry phase of the system when complex eigenvalues start to appear, leading to exponentially growing/decreasing behavior. Besides its great impact on the foundations of quantum theory~\cite{Znojil2008PRD,Gong2013JPA,Mostafazadeh2010IJGMMP,Schmidt2013RSTA,Deffner2015PRL,Brody2016JPA}, PT-symmetry is becoming increasingly important in optical physics and is playing a significant role in many intriguing applications such as single-mode lasing~\cite{Feng2014Science,Hodaei2014Science}, negative refraction~\cite{Fleury2014PRL,Monticone2016PRX}, nonreciprocal propagation~\cite{Peng2014NPhys,Chang2014Nphoton} and so on.

We show here that the QASER is indeed linked to a broken PT-symmetry phase. This is illustrated in Sec.~\ref{sec-basic-model} by firstly simplifying the basic model of QASER for two coupled oscillators at short evolution time when only the fundamental Floquet components are important, the resulting equations of motion are found to be equivalent to a two-mode PT-symmetric system as shown in Sec.~\ref{sec-connection-to-PT}. With the help of the two-mode PT-symmetric system, we are able to get the approximated quantum Langevin equation for the QASER in the short evolution time as given in Sec.~\ref{sec-Quantization}, finding that the fundamental Floquet components of QASER can be exponentially generated from quantum vacuum fluctuations.

\section{Basic model of QASER for two coupled oscillators}\label{sec-basic-model}

\begin{figure}[t]
\begin{center}
 \includegraphics[width=0.45\textwidth]{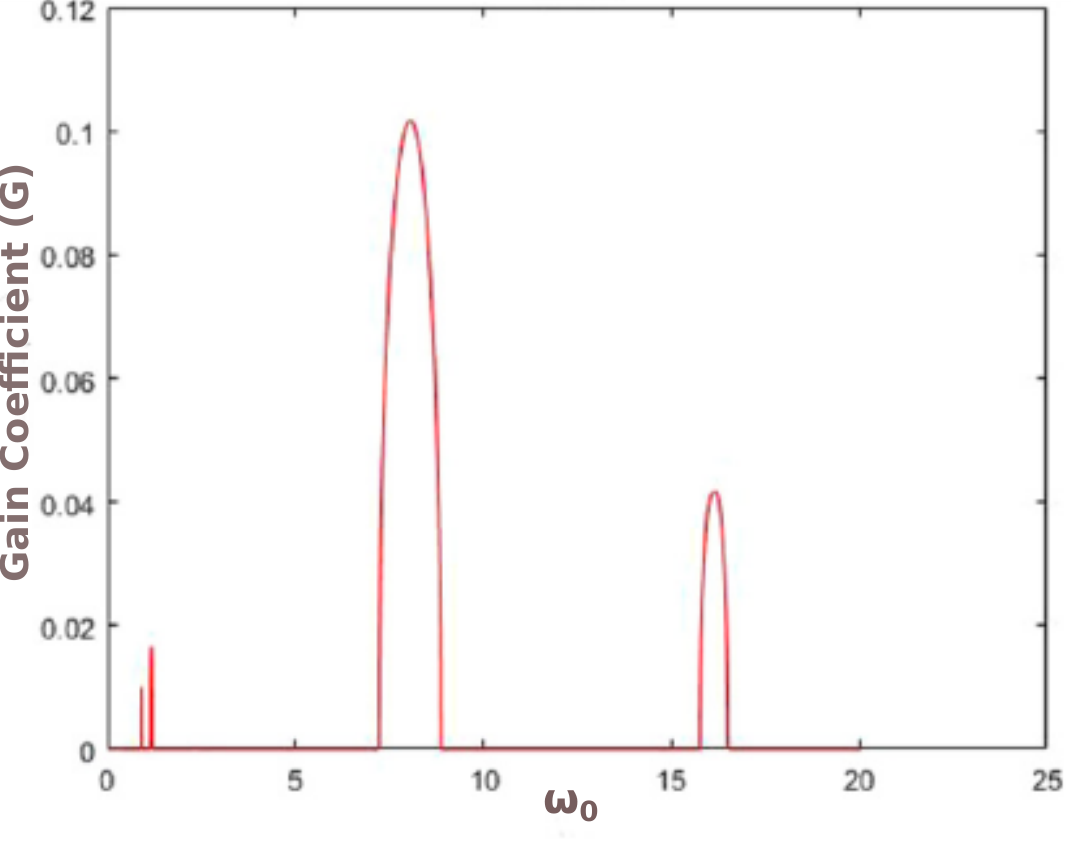}
 \caption{Gain coefficient $G$ as a function of $\omega_{0}$ for the two coupled oscillators. Parameters are chosen as before: $\Omega = \omega_0/2,~\delta= 0.4,\nu_{d}=2$. (Adopted from Ref.~\cite{Chen2016PS}, Fig. 2).}
\label{fig-gain}
 \end{center}
\end{figure}

We start from the basic model of QASER for two coupled oscillators~\cite{Svidzinsky2013PRX,Chen2016PS}
\begin{subequations}
\label{QASER-model}
\begin{align}
 &\ddot{\phi}_{1} + \omega^{2}_{0}\phi_{1} - \Omega^{2}\phi_{2}=0\,,\\[2ex]
 &\ddot{\phi}_{2} + \omega^{2}_{0}\phi_{2} - \Omega^{2}(1+\delta\cos \nu_{d} t)\phi_{1}=0\,.
\end{align}
\end{subequations}
Here $\omega_0$ is the frequency of the oscillators, $\omega_0\gg\Omega$ where $\Omega$ is the coupling constant between the two oscillators, and $\delta\ll1$ is the modulation amplitude of the low-frequency field. The coupled system has a pair of closely-lying normal-mode frequencies 
$\omega_{1}=\sqrt{\omega^{2}_{0}-\Omega^{2}}$ and 
$\omega_{2}=\sqrt{\omega^{2}_{0}+\Omega^{2}}$. Under the modulation of the low-frequency driving field $\nu_{d}\ll\omega_{0}$, the coupled system shows strong amplification for $\phi_{1}$ and $\phi_{2}$ with gain coefficient $G=\Omega^{2}\delta/(8\omega_{1}\omega_{2})$ when the modulation frequencies matches the frequency difference between the two normal modes, i.e., 
$\nu_{d} = \omega_{2}-\omega_{1}$
as shown in Fig.~\ref{fig-gain}. The coupled-oscillators model may find realizations in coupled PT symmetric microresonators~\cite{Peng2014NPhys,Chang2014Nphoton,Liu2016PRL,He2016PRA} or optomechanical systems~\cite{Xu2015PRA}. By introducing two new variables
\begin{subequations}
 \begin{align}
  X &= \phi_{1} + \phi_{2}\,,\\[2ex]
  Y &= \phi_{1} -\phi_{2}\,,
 \end{align}
\end{subequations}
Eqs.~(\ref{QASER-model}) are transformed into
\begin{subequations}
\label{QASER-model-XY}
 \begin{align}
  \ddot{X} + \omega^2_{1}X -\frac{1}{2}\Omega^{2}\delta\cos (\nu_{d} t) (X+Y) =0\,,\\[2ex]
  \ddot{Y} + \omega^2_{2}Y +\frac{1}{2}\Omega^{2}\delta\cos (\nu_{d} t) (X+Y) =0\,,
 \end{align}
\end{subequations}
In the above equations, all the variables  like $\phi_1, \phi_2,X$ and $Y$ are real functions of $t$. Now we will introduce two complex coordinates to further simplify the coupled equations
\begin{subequations}
\label{alpha-beta-definitions}
 \begin{align}
 \alpha &= (X+\frac{i}{\omega_{1}}\dot{X})e^{i\omega_{1}t}\,,\\[2ex]
 \beta &= (Y+\frac{i}{\omega_{2}}\dot{Y})e^{i\omega_{2}t}\,,
\end{align}
\end{subequations}
one can then obtain (see Appendix \ref{app-A} for more detail)
\begin{subequations}
 \begin{align}
  \dot{\alpha} &= \frac{i\Omega^2\delta}{4\omega_{1}}(\alpha e^{-i\omega_1 t} + \beta e^{-i\omega_2 t} + c.c.)\cos(\nu_{d} t)e^{i\omega_1 t}\,,\\[2ex]
  \dot{\beta} &= -\frac{i\Omega^2\delta}{4\omega_{2}}(\alpha e^{-i\omega_1 t} + \beta e^{-i\omega_2 t} + c.c.)\cos(\nu_{d} t)e^{i\omega_2 t}\,,
  \end{align}
\end{subequations}
where ``c.c.'' represents the counter-rotating terms and can be neglected. At the resonance condition $\omega_{2}-\omega_{1}=\nu_{d}$, according to the Floquet theory one can write the solution as 
\begin{subequations}
\label{alpha-beta-expansion}
 \begin{align}
  \alpha &= \sum_{n}\alpha_{n}e^{in\nu_{d} t}\,,\\[2ex]
  \beta &= \sum_{n}\beta_{n}e^{in\nu_{d} t}\,.
 \end{align}
\end{subequations}
The full Floquet analysis is given in Ref.~\cite{Chen2016PS}. When considering the evolution of the system for a short time or for small coupling constant $\delta$, only the base component ($n=0$) is important, resulting in (see Appendix \ref{app-A} for more detail)
\begin{subequations}
\label{QASER-base}
 \begin{align}
  \dot{\alpha}_{0} = \frac{i\Omega^2\delta}{8\omega_{1}}\beta_{0}\,,\\[2ex]
  \dot{\beta}_{0} = -\frac{i\Omega^2\delta}{8\omega_{2}}\alpha_{0}\,,
 \end{align}
\end{subequations}
Eqs.~(\ref{QASER-base}) would also now serve as the starting point that we discuss the relation between the QASER and PT symmetric system in the next section. Furthermore, we can also find from Eqs.~(\ref{QASER-base}) that 
\begin{align}
 \ddot{\alpha}_{0} = \frac{\Omega^4\delta^2}{64\omega_{1}\omega_{2}}\alpha_{0}\,.
\end{align}
Thus the gain coefficient for $\alpha_{0}$ is calculated as
\begin{align}
 G_{0}=\frac{\Omega^2\delta}{8\sqrt{\omega_{1}\omega_{2}}}\,.
\end{align}
Choosing the parameters as given in Fig.~\ref{fig-gain} and $\omega_0\simeq 8$ it is found that $G_{0}\simeq 0.1$ which agrees perfectly well with the numerical results as shown in Fig.~\ref{fig-gain}.

\section{Hidden PT symmetry of QASER equations} \label{sec-connection-to-PT}
In this section, we try to show the connection between the QASER and PT symmetric system. In order to illustrate this, we first introduce two transformations as follows 
\begin{subequations}
\label{PT-transform}
 \begin{align}
  &a = \frac{\alpha_{0} + i\beta_{0}}{\sqrt{2}}\,, &b = \frac{\alpha_{0} - i\beta_{0}}{\sqrt{2}}\,,\\[2ex]
  &\frac{\Omega^2\delta}{8\omega_{1}} = J + g\,, &\frac{\Omega^2\delta}{8\omega_{2}} = g - J\,,
 \end{align}
\end{subequations}
Thus Eqs.~(\ref{QASER-base}) reduce to 
\begin{subequations}
\label{PT-classical}
 \begin{align}
  \dot{a} &= g a - J b\,,\\[2ex]
  \dot{b} &= J a - g b\,.
 \end{align}
\end{subequations}
It can be seen that $a$ has linear gain $g$ $(\delta>0)$ and $b$ experiences balanced linear loss simultaneously, and the parameter $J$ gives the coupling between the two modes $a$ and $b$. This equation can be rewritten in matrix form
\begin{align}
\label{PT-symmetric-Qaser}
 \dot{\textbf{v}}=i\hat{H}_{\text{eff}}\textbf{v}\,,
\end{align}
where $\textbf{v}=\{a,b\}^{T}$ is the vector of variables, and $\hat{H}_{\text{eff}}$ can be considered as the effective Hamiltonian of the system whose form is 
\begin{align}
 \hat{H}_{\text{eff}}=\begin{bmatrix}
                       -ig & iJ\\[2ex]
                       -iJ & ig
                      \end{bmatrix}\,.
\end{align}
It can be easily seen that $\hat{H}_{\text{eff}}$ is not Hermitian since $\hat{H}_{\text{eff}}=\hat{H}^{\dagger}_{\text{eff}}$ unless $g=0$. One can prove that $\hat{H}_{\text{eff}}$ is invariant under the combination of time reversal $\hat{T}:i\rightarrow-i$ and parity reflection $\hat{P}$ which is defined as
\begin{align}
 \hat{P}=\begin{bmatrix}
                       0 &~~ 1\\[2ex]
                       1 &~~0
                      \end{bmatrix}\,,
\end{align}
which means $\hat{H}_{\text{eff}}$ is PT-symmetric. The eigenvalues of $\hat{H}_{\text{eff}}$ can be found as
\begin{align}
 \lambda_{\pm} &= \pm i\lambda\,,
\end{align}
where $\lambda=\sqrt{g^2-J^{2}}=G_0>0$ is real, thus the two eigenvalues are purely imaginary. It means that the system is in the fully broken PT-symmetry phase, resulting in exponentially growth for the two modes $a$ and $b$, and thus $\alpha_0$ and $\beta_0$.

We have seen that the QASER is equivalent to a two-mode PT-symmetric system when only considering the fundamental Floquet components, a natural question one might ask would be whether or not this relation still holds when extending to higher Floquet modes? A short answer is probably no. This can be seen from the coupled equations for the Floquet components after the rotating-wave approximation as given by Eqs.~(\ref{Floquet-series-eqs}), where the $n-2$ and $n+2$ terms (i.e., $\alpha_{n-2}$ and $\beta_{n+2}$) are asymmetrically coupled in the sense they only appear in one of the $n$-th equations (either in $a_{n}$ or in $b_{n}$), and thus introduce additional terms after the transformations as defined in Eqs.~(\ref{PT-transform}) for all $n$ components, the resulting effective Hamiltonian when including higher Floquet modes is then not PT-symmetric.

\section{Quantization\label{sec-Quantization}}
Having realized the PT-symmetric nature of the QASER equations, we proceed to quantize them in order to study the quantum nature of the system. 
Let's first have a closer look at the definitions of $\alpha$ and $\beta$ given in Eqs.~(\ref{alpha-beta-definitions}) and (\ref{alpha-beta-expansion}), it is apparent that $\alpha$ and $\beta$ are reminiscent of multi-mode annihilation operators, and $\alpha_{0}$ and $\beta_{0}$ are similar to single-mode operators, as well as $a$ and $b$. We may then replace $a$ and $b$ by $\hat{a}$ and $\hat{b}$ respectively in order to quantize Eqs.~(\ref{PT-classical}). However, this is not enough to get the quantum equations owing to the linear gain and loss which would introduce quantum noises~\cite{Scully-book,Schomerus2010PRL,Yoo2011PRA,Agarwal2013PRA}. After including the random noise terms, one can obtain the quantum Langevin equations for $\hat{a}$ and $\hat{b}$
\begin{subequations}
\label{PT-Quantum}
 \begin{align}
  \frac{d\hat{a}}{dt} &= g \hat{a} - J \hat{b} + \hat{f}_{a}\,,\\[2ex]
  \frac{d\hat{b}}{dt} &= J \hat{a} - g \hat{b} + \hat{f}_{b}\,,
 \end{align}
\end{subequations}
where $\hat{f}_{a}$ and $\hat{f}_{b}$ represent the quantum noise for the two modes respectively, and satisfy the correlations
\begin{subequations}
 \begin{align}
  \langle\hat{f}^{\dagger}_{a}(t)\hat{f}_{a}(t^{'})\rangle &= 2g\delta(t-t^{'}), &\langle\hat{f}_{a}(t)\hat{f}^{\dagger}_{a}(t^{'})\rangle &= 0\,,\\[2ex]
  \langle\hat{f}_{b}(t)\hat{f}^{\dagger}_{b}(t^{'})\rangle &= 2g\delta(t-t^{'}), &\langle\hat{f}^{\dagger}_{b}(t)\hat{f}_{b}(t^{'})\rangle &= 0\,.
 \end{align}
\end{subequations}
An exceptional feature of Eqs.~(\ref{PT-Quantum}) is that spontaneous generation grows exponentially from vacuum for PT-symmetry broken phase. This can be shown by calculating $\langle\hat{a}^{\dagger}\hat{a}\rangle$ and $\langle\hat{b}^{\dagger}\hat{b}\rangle$ when the input states for both fields are vacuum 
\begin{subequations}
 \begin{align}
\langle\hat{a}^{\dagger}\hat{a}\rangle &= \frac{J^{2}g}{\lambda^2}\left[ \frac{g}{J^{2}} (\cosh(2\lambda t)-1) + \frac{\lambda^2+g^2}{2\lambda J^{2}}\sinh(2\lambda t) - t\right],\\[2ex]
\langle\hat{b}^{\dagger}\hat{b}\rangle &=  \frac{J^{2}g}{\lambda^2}\left(\frac{\sinh(2\lambda t)}{2\lambda}-t\right)\,.
 \end{align}
\end{subequations}
For the values $J$ and $g$ defined in Eqs.~(\ref{PT-transform}), it has been shown in the previous section that PT symmetry is broken in the system, and exponential growth of spontaneous generation appears for $\hat{a}$ and $\hat{b}$ as shown above. 

One can then obtain the quantized form of Eqs.~(\ref{QASER-base}) by transforming $\hat{a}$ and $\hat{b}$ back to $\hat{\alpha}_{0}$ and $\hat{\beta}_{0}$
\begin{subequations}
 \begin{align}
\frac{d\hat{\alpha}_{0}}{dt} &= \frac{i\Omega^2\delta}{8\omega_{1}}\hat{\beta}_{0} + \hat{f}_{\alpha}\,,\\[2ex]
 \frac{d\hat{\beta}_{0}}{dt} &= -\frac{i\Omega^2\delta}{8\omega_{2}}\hat{\alpha}_{0} + \hat{f}_{\beta}\,,
 \end{align}
\end{subequations}
and the noise terms are given by
\begin{subequations}
 \begin{align}
 \hat{f}_{\alpha} = \frac{\hat{f}_{a}+\hat{f}_{b}}{\sqrt{2}}\,,\\
 \hat{f}_{\beta} = \frac{\hat{f}_{a}-\hat{f}_{b}}{\sqrt{2}i}\,.
\end{align}
\end{subequations}
Then the nonzero correlations between the quantum fluctuations can be derived
\begin{subequations}
 \begin{align}
  \langle\hat{f}^{\dagger}_{\alpha}(t)\hat{f}_{\alpha}(t^{'})\rangle &= \langle\hat{f}^{\dagger}_{\beta}(t)\hat{f}_{\beta}(t^{'})\rangle = \frac{(\omega_{1}+\omega_2)\Omega^2\delta}{16\omega_{1}\omega_2}\delta(t-t^{'})\,,\\[2ex]
  \langle\hat{f}_{\alpha}(t)\hat{f}^{\dagger}_{\alpha}(t^{'})\rangle &= \langle\hat{f}_{\beta}(t)\hat{f}^{\dagger}_{\beta}(t^{'})\rangle = \frac{(\omega_{1}+\omega_2)\Omega^2\delta}{16\omega_{1}\omega_2}\delta(t-t^{'})\,,\\[2ex]
  \langle\hat{f}^{\dagger}_{\alpha}(t)\hat{f}_{\beta}(t^{'})\rangle &= \langle\hat{f}_{\alpha}(t)\hat{f}^{\dagger}_{\beta}(t^{'})\rangle = -i\frac{(\omega_{1}+\omega_2)\Omega^2\delta}{16\omega_{1}\omega_2}\delta(t-t^{'})\,,\\[2ex]
  \langle\hat{f}^{\dagger}_{\beta}(t)\hat{f}_{\alpha}(t^{'})\rangle &= \langle\hat{f}_{\beta}(t)\hat{f}^{\dagger}_{\alpha}(t^{'})\rangle = i\frac{(\omega_{1}+\omega_2)\Omega^2\delta}{16\omega_{1}\omega_2}\delta(t-t^{'})\,.
 \end{align}
\end{subequations}

\begin{figure}[t!]
\begin{center}
 \includegraphics[width=0.45\textwidth]{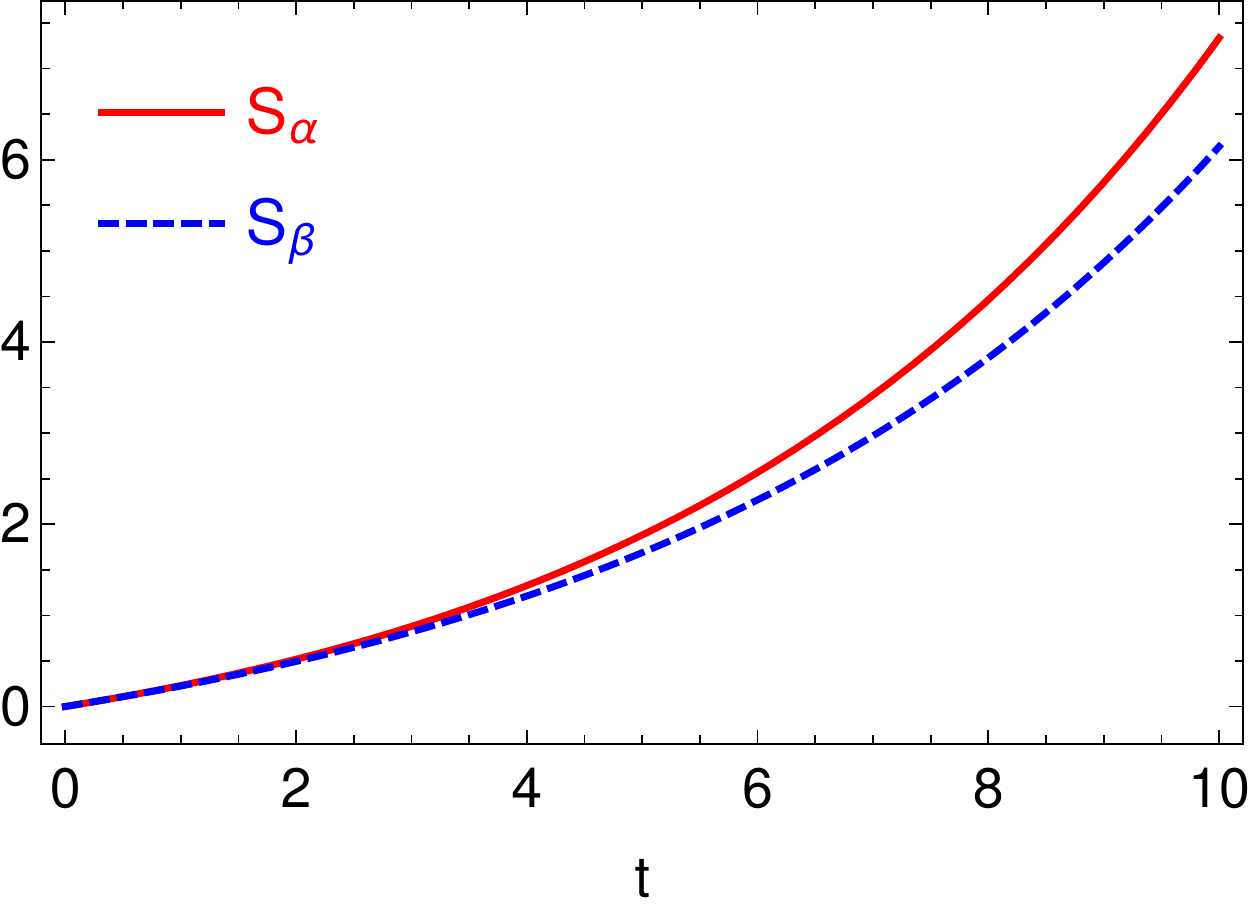}
 \caption{Exponential spontaneous generation from quantum vacuum fluctuations for the fundamental Floquet components of QASER. Parameters are chosen as before: $\Omega = \omega_0/2,~\delta= 0.4,~\omega_0=8$. }
 \label{VacuumGeneration}
 \end{center}
\end{figure}

One can then calculate the spontaneous generation for $\hat{\alpha}_{0}$ and $\hat{\beta}_{0}$ when the input state for both fields is vacuum
\begin{subequations}
 \begin{align}
S_{\alpha}&= \frac{\omega_1 +\omega_{2}}{16\omega^{2}_{1}\omega_2}\left[(\omega_{1}-\omega_2)\Omega^{2}\delta t + 8\omega_1\omega_2(\cosh (2\lambda t) -1)\right.\nonumber\\[2ex]
&\quad\quad\quad\quad\,\,\left.+4\sqrt{\omega_1\omega_2}(\omega_1+\omega_2)\sinh(2\lambda t)\right]\,,\\[2ex]
S_{\beta} &= \frac{\omega_1 +\omega_{2}}{16\omega_{1}\omega^{2}_2}\left[(\omega_{2}-\omega_1)\Omega^{2}\delta t + 8\omega_1\omega_2(\cosh (2\lambda t) -1)\right.\nonumber\\[2ex]
&\quad\quad\quad\quad\,\,\left.+4\sqrt{\omega_1\omega_2}(\omega_1+\omega_2)\sinh(2\lambda t)\right]\,.
 \end{align}
\end{subequations}
where $S_{O}=\langle\hat{O}^{\dagger}_{0}\hat{O}_{0}\rangle$ with $O\in\{\alpha,\beta\}$. For illustration, we have plot $S_{\alpha}$ and $S_{\beta}$ against evolution time $t$ as shown in Fig.~\ref{VacuumGeneration}. Similar to $\hat{a}$ and $\hat{b}$, $\hat{\alpha}_{0}$ and $\hat{\beta}_{0}$ show exponential growing behavior against time $t$ even when both of the input fields are vacuum. The quantum generation starting from vacuum is different from semiclassical calculations where one concentrates on gain by assuming a small seed field.

Although we adopted an open system approach to understand gain in the system; it is possible to develop an Hamiltonian framework~\cite{Giese2015} for the oscillator system described by Eq.~(\ref{QASER-model-XY}). In this work by Giese {\it et al} both the oscillator have time dependent frequency modulation in addition to a coupling which depends on the low frequency drive. The most important aspect of the Hamiltonian description is that one of the oscillators has a negative mass and inverted potential. The gain is then interpreted as arising from the oscillator with inverted energy spectrum. Giese {\it et al} also argue by considering a simpler coupled oscillator Hamiltonian that gain can only arise if one of the oscillators has inverted spectrum.

\section{Conclusions}

In conclusion, we have found that the coupled-oscillators model of QASER for short evolution time or weak parametric coupling when higher Floquet components are negligible is equivalent to a two-mode parity-time (PT) symmetric system. In the context of PT symmetry, we interpret the exponential growth in the QASER is associated to the PT symmetry breaking of the system which leads to imaginary eigenvalues. Moreover, by investigating the quantum nature of the two-mode PT symmetric system, we have derived the quantum Langevin equations for the QASER and demonstrate that the exponential gain originates from the quantum vacuum fluctuations.

G. S. A. thanks colleagues, especially Anatoly Svidzinsky for discussions and the Texas A \& M University for support.  We acknowledge the support of Office of Naval Research Grant N00014-16-1-3054 and Robert A. Welch Foundation Award A1261.

\appendix

\numberwithin{equation}{section}

\section{\label{app-A} Simplifications of the coupled equations}
Here we give the detailed steps on the derivation of Eq.~(\ref{QASER-base}). We first introduce two complex variables as
\begin{subequations}
 \begin{align}
  &z_{1} = X + \frac{i}{\omega_1}\dot{X}\,,\\[2ex]
  &z_{2} = Y + \frac{i}{\omega_2}\dot{Y}\,,
 \end{align}
\end{subequations}
then one can find that 
\begin{subequations}
 \begin{align}
  \dot{z}_{1} + i\omega_1 z_1 = \frac{i}{\omega_1}(\ddot{X}+ \omega^2_{1}X)\,,\\[2ex]
  \dot{z}_{2} + i\omega_2 z_2 = \frac{i}{\omega_2}(\ddot{Y}+ \omega^2_{2}Y)\,.
 \end{align}
\end{subequations}
Inserting these expressions into Eqs.~(\ref{QASER-model-XY}) leads to
\begin{subequations}
 \begin{align}
 \dot{z}_{1} + i\omega_{1}z_{1} - \frac{i\Omega^2\delta}{4\omega_{1}}\cos(\nu_{d} t)(z_{1}+z^{*}_{1}+z_{2}+z^{*}_{2})=0\,,\\[2ex]
 \dot{z}_{2} + i\omega_{2}z_{2} + \frac{i\Omega^2\delta}{4\omega_{2}}\cos(\nu_{d} t)(z_{1}+z^{*}_{1}+z_{2}+z^{*}_{2})=0\,.
 \end{align}
\end{subequations}
Now we define 
\begin{subequations}
 \begin{align}
  \alpha &= z_{1} e^{i\omega_1 t}\,,\\[2ex]
  \beta &= z_{2} e^{i\omega_2 t}\,,
 \end{align}
\end{subequations}
here $\alpha$ and $\beta$ represent the slow-varying parts of $z_{1}$ and $z_{2}$, respectively. Then one obtains
\begin{subequations}
\label{Floquet-system}
 \begin{align}
  \dot{\alpha} &= \frac{i\Omega^2\delta}{4\omega_{1}}(\alpha e^{-i\omega_1 t} + \beta e^{-i\omega_2 t} + c.c.)\cos(\nu_{d} t)e^{i\omega_1 t}\,,\\[2ex]
  \dot{\beta} &= -\frac{i\Omega^2\delta}{4\omega_{2}}(\alpha e^{-i\omega_1 t} + \beta e^{-i\omega_2 t} + c.c.)\cos(\nu_{d} t)e^{i\omega_2 t}\,,
  \end{align}
\end{subequations}
where ``c.c.'' denote the counter-rotating terms which we will neglect in the following. Furthermore we consider the resonant case when $\omega_{2}-\omega_{1}=\nu_{d}$, physically corresponding to the combination parametric resonance~\cite{Svidzinsky2013PRX,Chen2016PS}. Eqs.~(\ref{Floquet-system}) are a Floquet system whose solution can be written as
\begin{subequations}
 \begin{align}
  \alpha &= \sum_{n}\alpha_{n}e^{in\nu_{d} t}\,,\\[2ex]
  \beta &= \sum_{n}\beta_{n}e^{in\nu_{d} t}\,,
 \end{align}
\end{subequations}
Eqs.~(\ref{Floquet-system}) can be transformed into
\begin{subequations}
\label{Floquet-series-eqs}
 \begin{align}
  \dot{\alpha}_{n} = -in\nu_{d}\alpha_{n} + \frac{i\Omega^2\delta}{8\omega_{1}}(\alpha_{n-1}+\alpha_{n+1}+\beta_{n}+\beta_{n+2})\,,\\[2ex]
  \dot{\beta}_{n} = -in\nu_{d}\beta_{n} - \frac{i\Omega^2\delta}{8\omega_{2}}(\alpha_{n-2}+\alpha_{n}+\beta_{n-1}+\beta_{n+1})\,.
  \end{align}
\end{subequations}
We now consider the evolution of the system for a short time where the higher components in $\alpha_{n}$ and $\beta_{n}$ for $n\geq 1$ are negligible, Eqs.~(\ref{Floquet-series-eqs}) are simplified to
\begin{subequations}
 \begin{align}
  \dot{\alpha}_{0} = \frac{i\Omega^2\delta}{8\omega_{1}}\beta_{0}\,,\\[2ex]
  \dot{\beta}_{0} = -\frac{i\Omega^2\delta}{8\omega_{2}}\alpha_{0}\,.
 \end{align}
\end{subequations}

\bibliographystyle{apsrev4-1}
\bibliography{referencesbase}

\end{document}